\documentclass[12pt]{article}

\usepackage{graphics,amssymb,epsfig,float}
\usepackage[usenames,dvips]{color}
\usepackage{graphicx}% Include figure files
\usepackage{epsfig}% Include figure files
\usepackage{rotating}% Include figure files
\usepackage{dcolumn}% Align table columns on decimal point
\usepackage{bm}% bold math
\usepackage{cite}
\usepackage{amsmath}
\usepackage{array}
\usepackage{setspace}

\makeatletter

\@addtoreset{equation}{section}
\makeatother

\def\lsim{\;\raise0.3ex\hbox{$<$\kern-0.75em\raise-1.1ex\hbox{$\sim$}}\;}
\def\gsim{\;\raise0.3ex\hbox{$>$\kern-0.75em\raise-1.1ex\hbox{$\sim$}}\;}
\def\beq{\begin{equation}}   \def\eeq{\end{equation}}
\def\ba{\begin{array}}       \def\ea{\end{array}}
\def\bea{\begin{eqnarray}}   \def\eea{\end{eqnarray}}

\def\nl{\newline}

\begin{document}

\begin{titlepage}
\begin{flushright}
LPT Orsay 15-27
\end{flushright}

%\centerline{\bf \today}

\begin{center}

\begin{doublespace}

\vspace{1cm}
{\Large\bf Possible explanation of excess events in the search for
jets, missing transverse momentum and a $Z$ boson in $pp$ collisions} \\
\vspace{2cm}

{\bf{Ulrich Ellwanger$^a$}}\\
\vspace{1cm}
\it  $^a$ LPT, UMR 8627, CNRS, Universit\'e de Paris--Sud, 91405 Orsay,
France, and \\
\it School of Physics and Astronomy, University of Southampton,\\
\it Highfield, Southampton SO17 1BJ, UK

\end{doublespace}

\end{center}
\vspace{2cm}

\begin{abstract}
We study to which extent SUSY extensions of the Standard Model can describe the excess
of events of 3.0 standard deviations
observed by ATLAS in the on-$Z$ signal region, respecting
constraints by CMS on similar signal channels as well as constraints
from searches for jets and $E^{miss}_\text{T}$.
GMSB-like scenarios are typically in conflict with these constraints,
and do not reproduce well the shape of the $E^{miss}_\text{T}$ distribution
of the data. An alternative scenario with two massive neutralinos can
improve fits to the total number of events
as well as to the $H_\text{T}$ and $E^{miss}_\text{T}$ distributions.
Such a scenario can be realised within the NMSSM.
\end{abstract}

\end{titlepage}

\section{Introduction}
After the first run of the LHC at a center of mass (c.m.) energy of mostly
8~TeV, no significant excesses have been observed in searches for physics
beyond the Standard Model \cite{atlas_summary,cms_summary}.
These searches cover a wide range of possible
signatures, notably various combinations of jets, missing transverse energy
($E^{miss}_\text{T}$), $b$-jets and leptons (electrons or muons).

Same-flavour opposite-sign dileptons can be classified into ``off-$Z$''
leptons (typically with an
invariant mass $m_{ll} < 81$~GeV or $m_{ll} > 101$~GeV), and ``on-$Z$''
leptons with $81\ \text{GeV} < m_{ll} < 101$~GeV. Often, leptons and
in particular on-$Z$ dileptons are vetoed in order to suppress Standard
Model (SM) backgrounds. On the other hand, some decay cascades of
supersymmetric (SUSY) particles could be particularly rich in
off-$Z$ dileptons (in the presence of light sleptons), or on-$Z$ dileptons
if $Z$ bosons appear particularly frequently in these cascades.

Recently, results of searches for SUSY particles in events
with dileptons, jets and $E^{miss}_\text{T}$ have been published by the CMS
and ATLAS collaborations \cite{Khachatryan:2015lwa,Aad:2015wqa}.
The aim was to test scenarios of gluino pair production in which the
gluinos $\tilde{g}$ decay via sleptons (leading to off-$Z$ dileptons),
and scenarios of gauge mediated SUSY breaking (GMSB)
or generalised gauge mediation (GGM)  where the gluinos
undergo 3-body decays into quark pairs and a neutralino $\chi_1^0$.
The latter may decay subsequently
into a nearly massless gravitino $\tilde{G}$ and a $Z$~boson, leading to
on-$Z$ dileptons. The corresponding gluino decay chain is then
$\tilde{g} \to q+\bar{q}+\chi_1^0 \to q+\bar{q}+Z+\tilde{G}$.
Relevant parameters are the gluino mass $m_{\tilde{g}}$, the
neutralino mass $m_{\chi_1^0}$, and the branching fractions of the
involved decays.

Whereas no
significant excesses were observed by CMS in \cite{Khachatryan:2015lwa}
(up to an excess of 2.6~standard deviations in the dilepton mass window
$20\ \text{GeV} < m_{ll} < 70$~GeV), an excess of 3.0~standard deviations
was reported by ATLAS in \cite{Aad:2015wqa} in the on-$Z$ signal region:
Summing electron and muon pairs, 29 events passing the cuts were observed
versus $10.6 \pm 3.2$ background events expected. 
No attempt was made in \cite{Aad:2015wqa} to explain the excess in terms
of a specific model; instead, weaker exclusion limits than expected
were shown in the $m_{\tilde{g}}-m_{\chi_1^0}$ plane of GGM models.
Various studies of scenarios which could contribute to this excess have recently
been published \cite{Barenboim:2015afa,Vignaroli:2015ama,Allanach:2015xga,Kobakhidze:2015dra,
Cao:2015ara,Dobrescu:2015asa,Cahill-Rowley:2015cha,Lu:2015wwa,Liew:2015hsa}.

$Z$~bosons decay dominantly hadronically. Thus, whenever
gluinos are pair produced, in most cases each of
the two gluino cascades will produce no dileptons, but two hard
jets: either from $q+\bar{q}$ if $m_{\tilde{g}}\gg m_{\chi_1^0} \gsim M_Z$,
or from hadronic $Z$ decays if $m_{\tilde{g}}\gsim m_{\chi_1^0} \gg M_Z$
implying a neutralino much heavier than the gravitino, i.e.
energetic $Z$~bosons. Hence, both scenarios are subject to constraints
from ``standard'' searches for SUSY in events with hard jets and
$E^{miss}_{\text T}$ \cite{Aad:2014wea,Khachatryan:2015vra}, even if one
considers
simplified models where squarks are assumed to be decoupled and gluino
pair production is the only process taken into account. 

In order to study
the impact of these constraints on GMSB-like scenarios, we simulated
various configurations of gluino and $\chi_1^0$ masses. Using the latest
version 1.2.0 of CheckMATE \cite{Drees:2013wra} we found that constraints
from \cite{TheATLAScollaboration:2013fha} (a preliminary version of
\cite{Aad:2014wea}) on final states with jets and $E^{miss}_{\text T}$
are very restrictive, and
supersede even the recent CMS constraints from
\cite{Khachatryan:2015lwa} in the $m_{\tilde{g}}-m_{\chi_1^0}$ plane.
Exceptions are scenarios with reduced branching fractions for the
considered decay chain, without allowing for alternative final states
leading to jets and $E^{miss}_{\text T}$.

In the present paper we study to which extent a scenario with {\it two heavy}
neutralinos in the gluino decay cascade can contribute to the ATLAS
signal region, circumventing constraints from
searches for jets and $E^{miss}_{\text T}$.
The gluino decay cascade considered subsequently is of the form
\beq\label{eq:1.1}
\tilde{g} \to q+\bar{q}+\chi_2^0 \to q+\bar{q}+Z+\chi_1^0
\eeq
with
\beq
m_{\chi_2^0} \lsim m_{\tilde{g}},\qquad m_{\chi_1^0} \sim m_{\chi_2^0}
-100\;\text{GeV}
\eeq
and sketched in Fig.~1.

\begin{figure}[ht!]
\begin{center}
\hspace*{-6mm}
\psfig{file=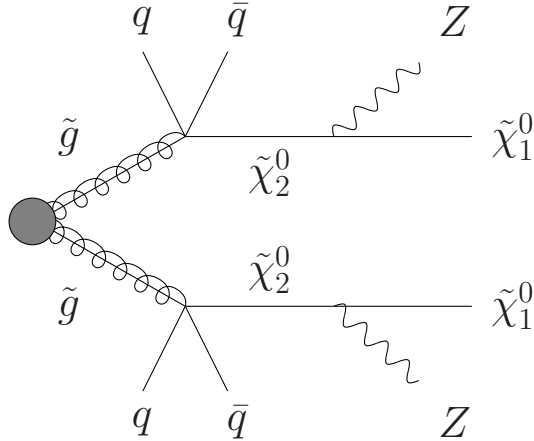, scale=0.80}
\end{center}
\caption{Gluino decay cascades involving two neutralinos $\chi_2^0$
and $\chi_1^0$.}
\label{fig:1}
\end{figure}

Now jets from both steps of the gluino decay cascade (including the jets from
the $Z$~boson) are relatively soft,
and constraints from searches for jets and $E^{miss}_{\text T}$ are
easier to satisfy unless the mass splitting $m_{\tilde{g}} - m_{\chi_2^0}$
is too large. Such a scenario has been considered recently also in
\cite{Cao:2015ara}. We will compare their results to ours in the conclusions.

In the following we consider first simplified models with
100\% branching fractions for both steps of the gluino decay cascade.
We simulated corresponding events, verified which scenarios satisfy 
the constraints from the CMS \cite{Khachatryan:2015lwa}
and other SUSY searches, and applied the cuts of ATLAS \cite{Aad:2015wqa}.
We will compare the signal rates and various distributions to the data
given in \cite{Aad:2015wqa}, and to a GMSB-like scenario (the latter with reduced
branching fractions in order to comply with constraints). Constraints from CMS
\cite{Khachatryan:2015lwa} prevent an excess as large as 3.0 standard
deviations in the ATLAS
signal region, but about 14 signal events on top of the background are
possible.

However, the question arises in which SUSY scenario such a neutralino
spectrum and, notably, such a dominant gluino decay cascade are possible: What can prevent
a dominant $\tilde{g} \to q+\bar{q}+\chi_1^0$ decay which is favored by phase
space? In GMSB the r\^ole of $\chi_1^0$ is played by the nearly
massless gravitino, which has tiny couplings to the MSSM-like sparticles
and is not produced unless, due to R-parity conservation, it is the only decay channel.
A heavier neutralino $\chi_1^0$ with small couplings to the MSSM-like sparticles,
as required in the present scenario, is
possible in the NMSSM \cite{Ellwanger:2009dp} in the form of the
singlino, the fermionic partner of the singlet superfield $S$ whose vacuum
expectation value generates dynamically a $\mu$-term (a SUSY mass term
for the two Higgs doublets in the MSSM) of the order of the SUSY breaking scale.
We find that there exist indeed scenarios within the parameter space of
the NMSSM for which the gluino decay cascade in Eq.~(\ref{eq:1.1}) is dominant.

In the next section we describe details of the simulation and cuts.
Results for simplified models and the description of a NMSSM scenario are
given in section 3. We conclude in section 4.

\section{Simulations and cuts}

We have simulated events at the LHC at 8~TeV using
MadGraph/\-MadEvent~\cite{Alwall:2011uj} which
includes Pythia~6.4 \cite{Sjostrand:2006za} for showering and
hadronisation. The emission of one additional hard jet was allowed in the
simulation in order to obtain realistic distributions for kinematical variables.
The production cross sections were obtained by
Prospino at NLO \cite{Beenakker:1996ch,Beenakker:1996ed}. 

First, the output was given to CheckMATE version 1.2.0 \cite{Drees:2013wra} which includes
the detector simulation DELPHES \cite{deFavereau:2013fsa} and compares
the signal rates to constraints from various search channels of ATLAS and
CMS. All searches present in CheckMATE version 1.2.0 have been verified;
the most relevant ones (with the largest ratio for the event yield to
$S^{95}_\text{obs}$ where $S^{95}_\text{obs}$ is the observed 95\% CL
upper bound)
are obtained from the ATLAS search for jets and $E^{miss}_{\text T}$ in
\cite{TheATLAScollaboration:2013fha}.

Second, the Pythia output was given directly to DELPHES and analysed
according to the object identification and selection criteria given
in \cite{Khachatryan:2015lwa,Aad:2015wqa}, respectively, and finally
the corresponding cuts were applied.

For the ATLAS on-$Z$ searches \cite{Aad:2015wqa} these were as follows:
$E^{miss}_\text{T} > 225$~GeV;  $\geq 2$ jets with $p_\text{T} > 35$~GeV;
two same-flavour opposite-sign leptons with $p_\text{T} > 25$~GeV for
the leading, $p_\text{T} > 10$~GeV for the sub-leading lepton;
$H_\text{T} > 600$~GeV where $H_\text{T} = p_\text{T}^\text{lepton,1}
+ p_\text{T}^\text{lepton,2} + \sum_i p_\text{T}^\text{jet,i}$ (including
jets with $p_\text{T} > 35$~GeV); and finally $81\ \text{GeV} < m_{ll} < 101$~GeV.
29 events passing the cuts were observed, whereas $10.6 \pm 3.2$
background events were expected. 

For the events passing the cuts, distributions
of $m_{ll}$, $E^{miss}_\text{T}$, $H_\text{T}$ and the jet multiplicity
$N_{\text{jets}}$
were shown in \cite{Aad:2015wqa} separately for the electron and muon channels.
These distributions were compared with those expected from two
GGM benchmark points with gluino masses and
neutralino masses of $(m_{\tilde{g}},m_{\chi_1^0})= (700,200)$\,GeV,
$(900,600)$\,GeV, respectively. We found, however, that both points
violate constraints from \cite{TheATLAScollaboration:2013fha} on final
states with jets and $E^{miss}_\text{T}$.

We compared the expected properties of the two GGM benchmark points in
\cite{Aad:2015wqa}
to the results of our simulation and found that they agree within $\sim 30\%$.
We conclude that the results of our simulations deviate by a systematic error
of up to $\sim 30\%$ from the more realistic (detector-) simulation of
the experimental collaboration. We can expect that this systematic
error cancels to a large extent
when comparing the properties of different simulated scenarios, but should be
taken into account when comparing to the actual data from \cite{Aad:2015wqa}.
Since it is of the same order (actually somewhat larger) than the difference
in the acceptances of dielectrons and dimuons in \cite{Aad:2015wqa}, we found it
reasonable to consider the sum of the data of dielectron and dimuon events
not only for the signal rate, but also for the kinematical distributions
and the expected SM background in order to obtain a larger statistics.

In the CMS on-$Z$ searches \cite{Khachatryan:2015lwa}, no cuts on $H_\text{T}$
were applied. Signal
jets were required to have $p_\text{T} > 40$~GeV. Six $E^{miss}_{\text T}$-
and $N_{\text{jets}}$-dependent on-$Z$ signal regions were defined: 
$E^{miss}_{\text T}=100-200,\ 200-300,\ >300$~GeV
and $N_{\text{jets}} \geq 2,\ \geq 3$, respectively. Finally the CMS and
ATLAS analyses differ slightly in the jet algorithms and in the lepton
acceptances. Comparing the signal rates obtained by our simulations of the two GMSB-like
benchmark points to the simulations in \cite{Khachatryan:2015lwa}
we found again that they agree within $\sim 30\%$.

As already stated above, no significant excesses were observed
in the on-$Z$ signal regions by CMS. Hence the event yields in the six
on-$Z$ signal regions lead to constraints on any scenarios which attempt to
explain the ATLAS excess. In the next section we discuss by means of benchmark points
to which extent the ATLAS excess can be matched in consideration of these constraints,
as well as constraints from \cite{TheATLAScollaboration:2013fha} on final
states with jets and $E^{miss}_{\text T}$.

\section{Results}

First we considered GMSB-like simplified models with a branching
fraction of 100\% for the 
$\tilde{g} \to q+\bar{q}+\chi_1^0 \to q+\bar{q}+Z+\tilde{G}$ decay
chain. Then, however, constraints from the search for jets and 
$E^{miss}_{\text T}$ in \cite{TheATLAScollaboration:2013fha} as tested
by CheckMATE \cite{Drees:2013wra} require $m_{\tilde{g}} \gsim 1050$~GeV
for small $m_{\chi_1^0} \sim 150$~GeV, and larger gluino masses for
larger $m_{\chi_1^0}$. Accordingly contributions to the ATLAS signal
region cannot exceed $\sim 5$ events. Moreover the distribution of
$E^{miss}_{\text T}$ and
$H_\text{T}$ peak towards large values (most events have
$H_\text{T}>1500$~GeV) in sharp contrast to the data in \cite{Aad:2015wqa}.

In realistic models, the branching fractions for the steps of the above
gluino decay chain can well be below 100\%. Below we will consider a
GMSB-like benchmark point ``GMSB'' with
$(m_{\tilde{g}},m_{\chi_1^0})= (800,600)$~GeV and a branching
fraction of 10\%  for the above decay chain; such a small branching fraction
makes it compatible with the CMS constraints. (A heavy
$\chi_1^0$ was chosen in order to shift the peak of the $H_\text{T}$
distribution towards lower values.) For the remaining 90\% of
the gluino decays one has to expect that, depending on the complete
spectrum and branching fractions, they contribute to the signal regions
in the search for jets and  $E^{miss}_{\text T}$ in
\cite{TheATLAScollaboration:2013fha}. One can make the somewhat optimistic
assumption that these contributions do not exceed 50\% of the contributions
of the $\tilde{g} \to q+\bar{q}+\chi_1^0 \to q+\bar{q}+Z+\tilde{G}$ decay
chain. Then this point remains within the constraints from
\cite{TheATLAScollaboration:2013fha}, but contributes about 10~events to
the ATLAS on-Z signal region.

Next we consider simplified models with two heavy neutralinos whose decay
chain is depicted in
Fig.~1. Assuming a branching fraction of 100\% for this decay chain,
gluinos can be as light as 800~GeV without conflict with
constraints from the search for jets and $E^{miss}_{\text T}$ in
\cite{TheATLAScollaboration:2013fha} -- under the condition, however,
that $m_{\chi_2^0}$ and $\ m_{\chi_1^0}$ are relatively large 
such that all jets remain relatively soft. We
studied two benchmark points P1 and P2 with
$(m_{\tilde{g}},m_{\chi_2^0},m_{\chi_1^0})= (800,790,690)$~GeV and
$(m_{\tilde{g}},m_{\chi_2^0},m_{\chi_1^0})= (800,600,500)$~GeV,
respectively. Such scenarios
belong to the few exceptions allowing for gluinos with a mass below
1~TeV, see the study in \cite{Evans:2013jna}.

For P1 with $m_{\tilde{g}} -
m_{\chi_2^0} = 10$~GeV the jets from the first step
$\tilde{g} \to q+\bar{q}+\chi_2^0$ of the decay cascade are very soft,
as are the jets from $Z$ decays from the second step
$\chi_2^0 \to Z+\chi_1^0$. Practically all energy of a single gluino
decay cascade goes into
$E^{miss}_{\text T}$. However, for typical kinematical configurations
the momenta of $\chi_1^0$ tend to be back-to-back in the transverse plane,
leading to a reduction of $E^{miss}_{\text T}$ of the complete event.
Only for relatively rare kinematical configurations (and/or extra jets
from initial state radiation as included in our simulation), 
$E^{miss}_{\text T}$ of the complete event can assume large values.
For P2 with $m_{\tilde{g}} -
m_{\chi_2^0} = 200$~GeV the jets from the first step
$\tilde{g} \to q+\bar{q}+\chi_2^0$ of the decay cascade are
harder, leading to less $E^{miss}_{\text T}$. One aim is to study the
impact of this difference on the distributions of kinematical
variables.

For all benchmark points we assumed practically
decoupled squarks with masses of 3~TeV; then the gluino pair production cross
section from prospino at NLO is 128~fb. (Since stops and sbottoms are assumed
to have masses of 3~TeV as well their pair production does not contribute to the signal.)
We deliberately chose identical gluino masses for all points in order to
maintain a common production cross section; therefore all differences in
contributions to signal regions and kinematical distributions originate from
the neutralino sector. The masses of the latter are recalled in Table~1 below.

\begin{table}[b!]
\begin{center}
\begin{tabular}{|c|c|c|c|c|}
\hline
 & & GMSB & P1 & P2   \\
\hline
Gluino/neutralino masses & &  &  &    \\
\hline
$m_{\tilde{g}}$& & 800 & 800 & 800  \\
\hline
$m_{\chi_1^0}$ (GMSB), $m_{\chi_2^0}$ (P1, P2)& & 600 & 790 & 600  \\
\hline
$m_{\tilde{G}}$ (GMSB), $m_{\chi_1^0}$ (P1, P2)& & 0 & 690 & 500  \\
\hline
\hline
Constraining signal regions & $S^{95}_\text{obs}$  &   &  &  \\
\hline
CMS, $N_{\text jets} \geq 2$, $100<E^{miss}_{\text T}< 200$
& 207 & 2.0 $\pm$ 0.6 	& 19.4 $\pm$ 5.8  & 55.7 $\pm$ 16.7  \\
\hline
CMS, $N_{\text jets} \geq 2$, $200<E^{miss}_{\text T}< 300$
& 20 & 2.6 $\pm$ 0.78 & 8.1 $\pm$ 2.4 &23.7 $\pm$ 7.1 \\
\hline
CMS, $N_{\text jets} \geq 2$, $300<E^{miss}_{\text T}$
 & 7.6 & 7.0 $\pm$ 2.1 & 6.1 $\pm$ 1.8 & 7.0 $\pm$ 2.1 \\
\hline
CMS, $N_{\text jets} \geq 3$, $100<E^{miss}_{\text T}< 200$
 & 89 & 1.9 $\pm$ 0.57 & 8.5 $\pm$ 2.6 & 48.4 $\pm$ 14.5 \\
\hline
CMS, $N_{\text jets} \geq 3$, $200<E^{miss}_{\text T}< 300$
 & 16.1 & 2.4 $\pm$ 0.72 & 4.7 $\pm$ 1.4 & 21.1 $\pm$ 6.3 \\
\hline
CMS, $N_{\text jets} \geq 3$, $300<E^{miss}_{\text T}$
 & 8 & 6.4 $\pm$ 1.9 & 3.8 $\pm$ 1.1 & 6.7 $\pm$ 2.0 \\
\hline
ATLAS, CT
 & 2.4 & 0.73 $\pm$ 0.22 & 0.79 $\pm$ 0.24 & 1.31 $\pm$ 0.39  \\
\hline
ATLAS, EM
 & 28.6 & 21.7 $\pm$ 6.5 & 1.32 $\pm$ 0.39 & 15.8 $\pm$ 4.7  \\
\hline
ATLAS, ET
 & 8.3 & 7.8 $\pm$ 2.3 & 0.70 $\pm$ 0.21 & 4.13 $\pm$ 1.24  \\
\hline
\hline
ATLAS on-$Z$ SR (obs. excess 18.4) & &9.8 $\pm$ 2.9& 5.6 $\pm$ 1.7& 13.6 $\pm$ 4.1 \\
\hline
\end{tabular}
\caption{Sparticle masses of the benchmark points GMSB,
P1 and P2 (in GeV), event yields including 30\% systematic errors from the simulation
of the benchmark points GMSB, P1 and P2 in the six signal regions of the CMS on-$Z$
searches  in \cite{Khachatryan:2015lwa}, and in the most constraining signal regions
CT, EM and ET of the ATLAS search \cite{TheATLAScollaboration:2013fha}.
The ranges of $E^{miss}_{\text T}$ for the six CMS signal regions are given in GeV.
The last line indicates the contributions to the ATLAS on-$Z$ signal region.}
\end{center}
\end{table}

In addition we indicate in the Table~1 in how far the benchmark points GMSB,
P1 and P2 satisfy constraints from the six signal regions of the CMS on-$Z$ searches 
in \cite{Khachatryan:2015lwa} (including 30\% systematic errors from the
simulation). The 95\%~CL upper limits for the six signal regions of the CMS on-$Z$
searches had already been obtained in \cite{Cao:2015ara}.
We find that the central values of event yields of the benchmark points
are below these 95\%~CL upper limits with the exception of P2 in the bins
$N_{\text jets} \geq 2$, $200<E^{miss}_{\text T}< 300$ and
$N_{\text jets} \geq 3$, $200<E^{miss}_{\text T}< 300$. However, taking the 
systematic errors from the simulation into account, the $CL_s=CL_{s+b}/CL_b$
values for P2 in these bins are 0.11 and 0.09, respectively, i.e. well above the
95\%~CL exclusion limit of 0.05.

Out of the 10 signal regions
in the ATLAS search \cite{TheATLAScollaboration:2013fha} for jets and
$E^{miss}_{\text T}$ we show the event yields for the signal regions
CT, EM and ET which give the largest ratio event yield/$S^{95}_\text{obs}$
for the points P1, P2 and GMSB, respectively. All these signal regions require
$E^{miss}_{\text T} > 160$~GeV, $p_T > 130$~GeV for the leading jet, and
$p_T > 60$~GeV for 3 additional jets (CT), $p_T > 60$~GeV for 5 additional
jets (EM and ET). EM and ET differ by $m_\text{eff}(incl.) > 1200/1500$~GeV,
respectively (see \cite{TheATLAScollaboration:2013fha} for more details).

We recall that the event yields for the
point GMSB assume only a branching fraction of 10\% into the considered
gluino decay chain. In Table~1, only the contributions from the simulated decay
chain are shown. Within the systematic error bars, $50\%$ more
events from other gluino decays are allowed to contribute to the signal regions in the
ATLAS search \cite{TheATLAScollaboration:2013fha} for jets and
$E^{miss}_{\text T}$ in order to saturate the bound from the signal region ET.

Finally we compare the contributions of the benchmark points GMSB,
P1 and P2 to the ATLAS on-$Z$ signal region, summing dielectrons and
dimuons, in the last line of Table~1.

We see that a price has to be paid for the very compressed
gluino$-\chi_2^0-\chi_1^0$ spectrum in P1: Due to the softness of the
jets, not enough jets satisfy the cut $N_{\text{jets}} \geq 2$.
The GMSB point seems to do quite well, despite its
gluino branching fraction being reduced by a factor $\sim 1/10$.
The best fit is given by
P2 with its less compressed gluino$-\chi_2^0-\chi_1^0$ spectrum.

Next we consider the distributions of kinematical variables.
As stated above we combine the ATLAS dielectron and dimuon data (despite
the different acceptances) in order to enhance the visibility of
possible trends. We only show the (dominant)
statistical error of the data; we are not in a position to combine the
partially correlated systematic errors. In the figures below we show
the data with the expected SM background contribution subtracted, with the aim
to expose possible desirable features of signal contributions (see
\cite{Aad:2015wqa} for the error attributed to the expected background).

\begin{figure}[t!]
\begin{center}
\hspace*{-6mm}
\psfig{file=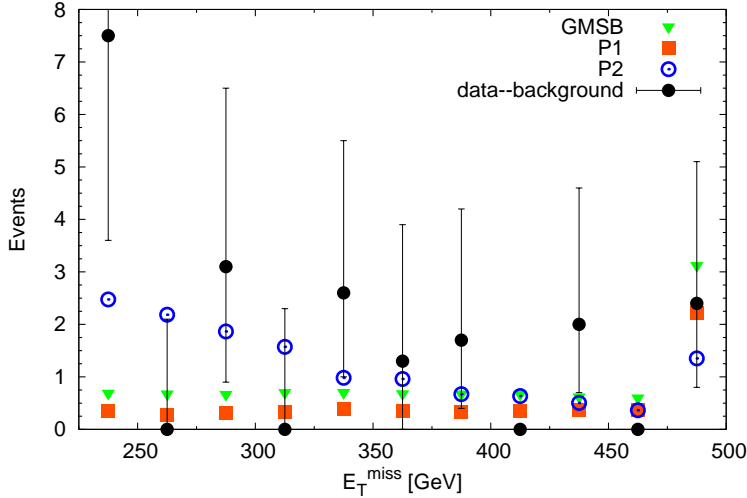, scale=0.80}
\end{center}
\caption{Comparison of $E^{miss}_\text{T}$ from the data in \cite{Aad:2015wqa}
(with the expected background subtracted) to the benchmark points GMSB, P1 and P2
defined in the text. Error bars on the data are statistical only. The rightmost
bin includes the overflow.}
\label{fig:2}
\end{figure}

We start with $E^{miss}_\text{T}$ in Fig.~2 where we compare the data with
the expected background subtracted to the GMSB scenario and with the two
heavy-neutralino benchmark points P1 and P2. 
We simulated 500.000 events for each scenario. Each expected event
for the LHC run~I as shown in Fig.~\ref{fig:2} corresponds to 10
simulated events, which allows to estimate the statistical errors. These
are smaller than the estimated systematic errors from our simulation,
and much smaller than the statistical error of the data.

The measured event numbers seem to decrease continuously with
$E^{miss}_\text{T}$ (within the error bars, and note that the rightmost
bin includes the overflow), whereas the
$E^{miss}_\text{T}$ distributions of the GMSB and P1 points are
nearly flat: In these scenarios, nearly all energy is
transformed into missing energy which prefers accordingly large values
of $E^{miss}_\text{T}$. (Note that $E^{miss}_{\text T}$ is shown after
the application of all cuts, notably on $H_T > 600$~GeV. For P1 with its
compressed spectrum this cut selects atypical kinematical configurations
with particularly large $E^{miss}_{\text T}$.)

For a quantitative comparison we compute the reduced $\chi^2$ statistic
\beq
\chi^2_\text{red} = \frac{1}{N_\text{bins}-1}
\sum_{i=1}^{N_\text{bins}}\frac{(N_{d-b}(i)-N_S(i))^2}{\sigma^2(i)}
\eeq
for each benchmark point, where $N_{d-b}(i)$ is the data with the expected
background subtracted (as shown in Fig.~2). $\sigma(i)$ combines the statistical
error of the data shown in Fig.~2 and the systematic error of 30\% of
our simulation (with respect to which the systematic error of the background is
negligible).

We obtained $\chi^2_\text{red} = 0.69$ for GMSB,
 $\chi^2_\text{red} = 0.85$ for P1 and  $\chi^2_\text{red} = 0.61$ for P2.
Hence the scenario P2 with its larger
splitting between the gluino and the $\chi_2^0$ masses describes
best the shape of the $E^{miss}_\text{T}$ distribution. Of course, the scenario P2
profits also from its larger total event rate.

\begin{figure}[t!]
\begin{center}
\hspace*{-6mm}
\psfig{file=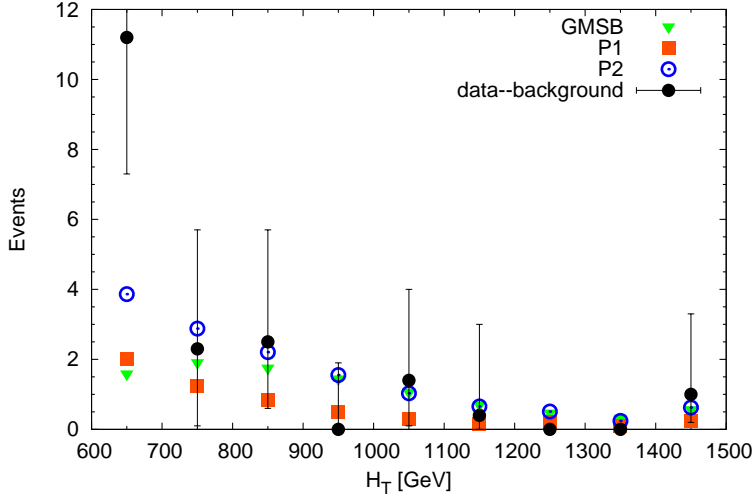, scale=0.80}
\end{center}
\caption{Comparison of $H_\text{T}$ from the data in \cite{Aad:2015wqa}
(with the expected background subtracted) to the benchmark points GMSB, P1 and P2
defined in the text. Error bars on the data are statistical only. The rightmost
bin includes the overflow.}
\label{fig:3}
\end{figure}

In Fig.~\ref{fig:3} we compare the data on $H_\text{T}$ (with the
expected background subtracted) with the GMSB scenario and the benchmark points
P1 and P2. Since $H_\text{T}$ represents most of the visible transverse
energy, the point P1 with its compressed spectrum peaks at low values
of $H_\text{T}$. This coincides with the trend of the data, but the
total signal rate (limited by constraints from CMS) is small, as indicated
in Table~1. 

For the reduced $\chi^2$ statistic we find 
$\chi^2_\text{red} = 0.54$ for GMSB,
$\chi^2_\text{red} = 0.69$ for P1 and  $\chi^2_\text{red} = 0.36$ for P2.
Again, the benchmark point P2 provides the best agreement with the shape of
the distribution despite its somewhat less compressed spectrum.

\begin{figure}[ht!]
\begin{center}
\hspace*{-6mm}
\psfig{file=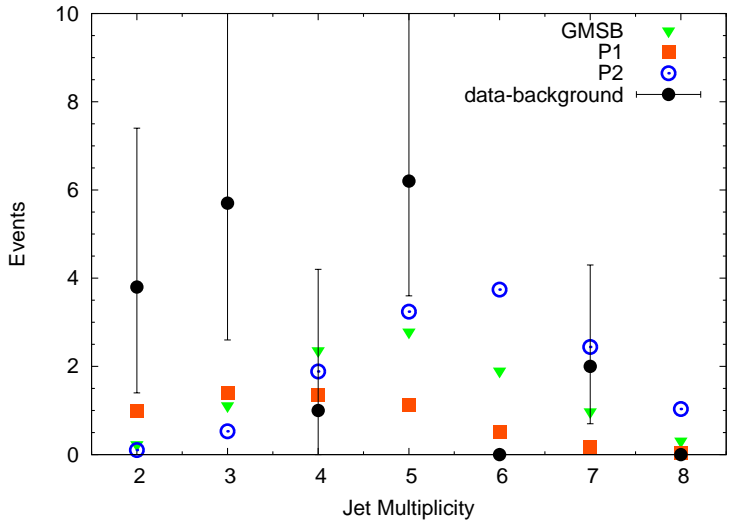, scale=0.80}
\end{center}
\caption{Comparison of the jet multiplicity from the data in \cite{Aad:2015wqa}
(with the expected background subtracted) to the benchmark points GMSB, P1 and P2
defined in the text. Error bars on the data are statistical only.}
\label{fig:4}
\end{figure}

Finally we turn to the distribution of the jet multiplicity in
Fig.~\ref{fig:4}. The trend of the data towards low jet multiplicities
is reproduced only by P1 with its excessively low signal rate. The jet multiplicity
of simulations is sensitive, amongst others, to the matching between
soft and hard QCD radiation, accordingly this quantity has to be considered
with some reserve.

For the reduced $\chi^2$ statistic we find 
$\chi^2_\text{red} = 1.03$ for GMSB,
$\chi^2_\text{red} = 1.08$ for P1 and  $\chi^2_\text{red} = 1.57$ for P2.
In this case the trend of the data is not
well reproduced by the point P2.
But since the scenario P2 provides the best fit to the ATLAS signal rate
and the $E^{miss}_\text{T}$ and $H_\text{T}$ distributions, it would be
interesting to know about SUSY extensions of the Standard Model which
share the features of this simplified model. As discussed in the
introduction, this is possible within the NMSSM. 

Using the spectrum
generator NMSSMTools~\cite{Ellwanger:2004xm,Ellwanger:2005dv} with
decay branching fractions computed by NMSDECAY \cite{Das:2011dg}
(based on HDECAY \cite{Djouadi:1997yw}) we found that the following
region of the parameter space of the $Z_3$-invariant NMSSM shares
the following properties with the point P2:
\begin{itemize}
\item Heavy (decoupled) squarks in order to satisfy constraints from
searches for events with jets and $E^{miss}_\text{T}$ in the presence
of a gluino with a mass of 800~GeV.

\item A bino-like neutralino $\chi_2^0$ with a mass of 600~GeV, but
winos and higgsinos slightly heavier than the gluino. (The
running gaugino masses do not satisfy SU(5)-like relations at the GUT
scale.)

\item A singlet-like neutralino $\chi_1^0$ with a mass of 500~GeV. Then
the branching fraction for the decay $\chi_2^0\to \chi_1^0 +Z$ is 100\%.

\item The loop-induced gluino two-body decay $\tilde{g} \to g + \chi_1^0$
should be suppressed, since it would not contribute to the signal.
It is induced by the higgsino component of
$\chi_1^0$, and can be of similar order of the desired gluino
three-body decay $\tilde{g} \to q + \bar{q} + \chi_2^0$.
The singlino-higgsino
mixing is proportional to the NMSSM-specific Yukawa coupling $\lambda$
\cite{Ellwanger:2009dp}, and $\lambda$ should not exceed $\sim 0.3$.
(The loop-induced gluino two-body decay $\tilde{g} \to g + \chi_2^0$
leads to similar signals as $\tilde{g} \to q + \bar{q} + \chi_2^0$,
but it can be expected that it would improve the jet multiplicity distribution
shifting it towards smaller values.)

\end{itemize}

The remaining parameters can be chosen to obtain a Standard Model-like
Higgs boson with a mass of $\sim 125$~GeV.
We have checked that a corresponding point in the parameter space with all
squark masses of 2.5~TeV (leading to a gluino pair production cross section of
$\sim 150$~fb in order to compensate for a gluino BR into $q + \bar{q}
+ \chi_2^0$ slightly below 100\%),
 $\tan\beta =3.75$, $\mu_{\text{eff}}=800$~GeV,
$\lambda \sim 0.28$, $\kappa \sim 0.087$,
$A_{\lambda} \sim 2.7$~TeV and $A_{\kappa} \sim -50$~GeV (see
\cite{Ellwanger:2009dp} for the definitions of the latter parameters)
has the properties of P2 and would not be distinguishable from P2 regarding
the different observables shown above in Figs.~\ref{fig:2}--\ref{fig:4}.

\section{Summary and conclusions}

We studied to which extent SUSY extensions of the SM can describe the excess
of events observed by ATLAS in the on-$Z$ signal region, respecting
constraints by CMS on similar signal channels as well as constraints
from searches for jets and $E^{miss}_\text{T}$. For viable scenarios
we compared the distribution of kinematical variables to the data, combining
dielectron and dimuon events.

Due to hadronic $Z$-decays, GMSB-like scenarios are typically in conflict
with constraints from searches for jets and $E^{miss}_\text{T}$.
Assuming a 100\% branching fraction for the gluino decay cascade including
the $\chi_1^0 \to \tilde{G} + Z$ decay, these scenarios become viable
only if the gluino has a mass above $\sim 1.05$~TeV
implying a small contribution ($<~5$ events) to the ATLAS signal region.
Reducing the branching fraction (including the $\chi_1^0 \to \tilde{G} + Z$ decay)
to $\sim 10\%$, lighter gluinos with $m_{\tilde{g}}\sim 800$~GeV
may contribute significantly to the signal region,
remaining within the 95\% CL limit of CMS. However, the $H_\text{T}$ and notably the
$E^{miss}_\text{T}$ distributions do not coincide well with the trends of the data. Therefore
we studied alternative scenarios with two massive neutralinos
$\chi_2^0$, $\chi_1^0$. 
In order to compare the impact of different neutralino spectra to GMSB-like
scenarios and among themselves for fixed gluino pair production cross sections
and gluino masses, we fixed the latter also to 800~GeV.

A very compressed $\tilde{g}-\chi_2^0-\chi_1^0$ spectrum reproduces
somewhat better the trend of the $H_\text{T}$ distribution, but does
not improve the shape of the $E^{miss}_\text{T}$ distribution.
In particular, the contribution to the ATLAS signal region cannot be
enhanced significantly while remaining within the 95\% CL limit of CMS.

A less compressed $\tilde{g}-\chi_2^0-\chi_1^0$ spectrum provides the
best fit to the total number of events in the ATLAS in the on-$Z$ signal region,
as well as to the $H_\text{T}$ and $E^{miss}_\text{T}$ distributions; only
the jet multiplicity is still not well reproduced.
(Larger $\tilde{g}-\chi_2^0$ mass splittings as assumed here become
again sensitive to constraints from searches for jets and $E^{miss}_\text{T}$.)
We found that such a scenario can be realised within the NMSSM.

A somewhat different approach has recently been persued in \cite{Cao:2015ara},
where the space of the two lightest neutralino masses within the NMSSM
was scanned systematically in order to maximise the contribution to the
ATLAS on-$Z$ signal region respecting existing constraints. The shapes of
the kinematical variables have not been studied, however. Still, their
main results coincide with ours: Whereas compressed spectra make it easier to
satisfy constraints from other SUSY searches, the contributions to the
ATLAS on-$Z$ signal region are suppressed as well. For gluino masses
below $\sim 800$~GeV, only a small corner in the plane of the two lightest
neutralino masses survives the 95\% CL limits of CMS. Within this corner
(for $\tilde{g}$, $\chi_2^0$, $\chi_1^0$ masses of 650, 565 and 465~GeV, respectively)
the authors found a maximal contribution of about 11 events to the ATLAS
on-$Z$ signal region.

Moreover, the authors of \cite{Cao:2015ara} considered constraints from
signal regions in \cite{Aad:2014wea} (2jW and 4jW) which are not implemented
in the CheckMATE version 1.2.0 \cite{Drees:2013wra} used here. The authors
applied these constraints to our benchmark point P2 and
obtained a ratio for the event yield/$S^{95}_\text{obs}$ of 1.19, i.e.
about 20\% too large, but within the systematic errors from the simulation.
A similar excess holds for this point actually also
for two CMS signal regions considered in Table~1. We recall that
identical gluino masses of 800~GeV were chosen for all points to simplify
comparisons. A slightly
heavier gluino mass of $\sim 825$~GeV would reduce the gluino pair production
cross section by $\sim 20\%$, but with little changes in the
decay kinematics if all mass splittings remain the same. Then this modified
point would pass all constraints without the help of systematic error bars,
but its contribution to the ATLAS on-$Z$ signal region would drop to 
$\sim 11$~events.
This number coincides with the maximum found in 
\cite{Cao:2015ara} for the slightly different point above.

Clearly, if the excess observed by ATLAS indicates the presence of
particles beyond the Standard Model, it should become more visible in
both ATLAS and CMS experiments at the run~II of the LHC. But since it
is present in the ATLAS analysis of the available data from run~I
we found it appropriate to discuss possible interpretations.

Within the class of models considered here, fits to the event numbers
and shapes of the ATLAS on-Z can be improved with respect to the GMSB
scenarios considered in \cite{Aad:2015wqa}. However, perfect fits would
lead to unacceptable tensions with constraints from other searches.

%\vfill

\section*{Acknowledgements}

We acknowledge support from European Union Initial
Training Network INVISIBLES (PITN-GA-2011-289442),
the ERC advanced grant Higgs@LHC, and from
the European Union Initial Training Network Higgs\-Tools
(PITN-GA-2012-316704).


\newpage

\begin{thebibliography}{99}

\bibitem{atlas_summary}
%\bibitem{Aad:2013wta}
  G.~Aad {\it et al.}  [ATLAS Collaboration],
  %``Search for new phenomena in final states with large jet multiplicities and missing transverse momentum at $\sqrt{s}$=8 TeV proton-proton collisions using the ATLAS experiment,''
  JHEP {\bf 1310} (2013) 130
   [Erratum-ibid.\  {\bf 1401} (2014) 109]
  [arXiv:1308.1841 [hep-ex]]; for an overview see
\nl {\sf 
https://twiki.cern.ch/twiki/bin/view/AtlasPublic/SupersymmetryPublicResults}

\bibitem{cms_summary} 
%\bibitem{Chatrchyan:2014lfa}
  S.~Chatrchyan {\it et al.}  [CMS Collaboration],
  %``Search for new physics in the multijet and missing transverse momentum final state in proton-proton collisions at $\sqrt{s}$= 8 TeV,''
  JHEP {\bf 1406} (2014) 055
  [arXiv:1402.4770 [hep-ex]]; for an overview see
\nl {\sf
https://twiki.cern.ch/twiki/bin/view/CMSPublic/PhysicsResultsSUS}

\bibitem{Khachatryan:2015lwa}
  V.~Khachatryan {\it et al.}  [CMS Collaboration],
  %``Search for physics beyond the standard model in events with two leptons, jets, and missing transverse momentum in pp collisions at sqrt(s) = 8 TeV,''
  JHEP {\bf 1504} (2015) 124
  [arXiv:1502.06031 [hep-ex]].

\bibitem{Aad:2015wqa}
  G.~Aad {\it et al.}  [ATLAS Collaboration],
  ``Search for supersymmetry in events containing a same-flavour opposite-sign dilepton pair, jets, and large missing transverse momentum in $\sqrt{s}=8$ TeV $pp$ collisions with the ATLAS detector,''
  arXiv:1503.03290 [hep-ex].


\bibitem{Barenboim:2015afa}
  G.~Barenboim, J.~Bernabeu, V.~A.~Mitsou, E.~Romero, E.~Torro and O.~Vives,
  ``METing SUSY on the Z peak,''
  arXiv:1503.04184 [hep-ph].

\bibitem{Vignaroli:2015ama}
  N.~Vignaroli,
  %``$Z$-peaked excess from heavy gluon decays to vectorlike quarks,''
  Phys.\ Rev.\ D {\bf 91} (2015) 11,  115009
  [arXiv:1504.01768 [hep-ph]].

\bibitem{Allanach:2015xga}
  B.~Allanach, A.~Raklev and A.~Kvellestad,
  %``Consistency of the recent ATLAS $Z+E_T^{\rm miss}$ excess in a simplified GGM model,''
  Phys.\ Rev.\ D {\bf 91} (2015) 095016
  [arXiv:1504.02752 [hep-ph]].

\bibitem{Kobakhidze:2015dra}
  A.~Kobakhidze, A.~Saavedra, L.~Wu and J.~M.~Yang,
  ``ATLAS Z-peaked excess in MSSM with a light sbottom or stop,''
  arXiv:1504.04390 [hep-ph].

\bibitem{Cao:2015ara}
  J.~Cao, L.~Shang, J.~M.~Yang and Y.~Zhang,
  %``Explanation of the ATLAS Z-Peaked Excess in the NMSSM,''
  JHEP {\bf 1506} (2015) 152
  [arXiv:1504.07869 [hep-ph]].

\bibitem{Dobrescu:2015asa}
  B.~A.~Dobrescu,
  ``Leptophobic boson signals with leptons, jets and missing energy,''
  arXiv:1506.04435 [hep-ph].

\bibitem{Cahill-Rowley:2015cha}
  M.~Cahill-Rowley, J.~L.~Hewett, A.~Ismail and T.~G.~Rizzo,
  ``The ATLAS Z + MET Excess in the MSSM,''
  arXiv:1506.05799 [hep-ph].

\bibitem{Lu:2015wwa}
  X.~Lu, S.~Shirai and T.~Terada,
  ``ATLAS Z Excess in Minimal Supersymmetric Standard Model,''
  arXiv:1506.07161 [hep-ph].

\bibitem{Liew:2015hsa}
  S.~P.~Liew, A.~Mariotti, K.~Mawatari, K.~Sakurai and M.~Vereecken,
  ``Z-peaked excess in goldstini scenarios,''
  arXiv:1506.08803 [hep-ph].

\bibitem{Aad:2014wea}
  G.~Aad {\it et al.}  [ATLAS Collaboration],
  %``Search for squarks and gluinos with the ATLAS detector in final states with jets and missing transverse momentum using $\sqrt{s}=8$ TeV proton--proton collision data,''
  JHEP {\bf 1409} (2014) 176
  [arXiv:1405.7875 [hep-ex]].

\bibitem{Khachatryan:2015vra}
  V.~Khachatryan {\it et al.}  [CMS Collaboration],
  %``Searches for supersymmetry using the M$_{T2}$ variable in hadronic events produced in pp collisions at 8 TeV,''
  JHEP {\bf 1505} (2015) 078
  [arXiv:1502.04358 [hep-ex]].

\bibitem{Drees:2013wra}
  M.~Drees, H.~Dreiner, D.~Schmeier, J.~Tattersall and J.~S.~Kim,
  %``CheckMATE: Confronting your Favourite New Physics Model with LHC Data,''
  Comput.\ Phys.\ Commun.\  {\bf 187} (2014) 227
  [arXiv:1312.2591 [hep-ph]].

\bibitem{TheATLAScollaboration:2013fha}
  The ATLAS collaboration,
  ``Search for squarks and gluinos with the ATLAS detector in final
states with jets and missing transverse momentum and 20.3 fb$^{-1}$ of
$\sqrt{s}=8$~TeV proton-proton collision data,''
  ATLAS-CONF-2013-047.

\bibitem{Ellwanger:2009dp}
  U.~Ellwanger, C.~Hugonie and A.~M.~Teixeira,
  %``The Next-to-Minimal Supersymmetric Standard Model,''
  Phys.\ Rept.\  {\bf 496} (2010) 1\newline
  [arXiv:0910.1785 [hep-ph]].


\bibitem{Alwall:2011uj}
  J.~Alwall, M.~Herquet, F.~Maltoni, O.~Mattelaer and T.~Stelzer,
  %``MadGraph 5 : Going Beyond,''
  JHEP {\bf 1106} (2011) 128
  [arXiv:1106.0522 [hep-ph]].

\bibitem{Sjostrand:2006za}
  T.~Sjostrand, S.~Mrenna and P.~Z.~Skands,
  %``PYTHIA 6.4 Physics and Manual,''
  JHEP {\bf 0605} (2006) 026
  [hep-ph/0603175].

\bibitem{Beenakker:1996ch}
  W.~Beenakker, R.~Hopker, M.~Spira and P.~M.~Zerwas,
  %``Squark and gluino production at hadron colliders,''
  Nucl.\ Phys.\ B {\bf 492} (1997) 51
  [hep-ph/9610490].

\bibitem{Beenakker:1996ed}
  W.~Beenakker, R.~Hopker and M.~Spira,
  ``PROSPINO: A Program for the production of supersymmetric particles in next-to-leading order QCD,''
  hep-ph/9611232.


\bibitem{deFavereau:2013fsa}
  J.~de Favereau {\it et al.}  [DELPHES 3 Collaboration],
  %``DELPHES 3, A modular framework for fast simulation of a generic collider experiment,''
  JHEP {\bf 1402} (2014) 057\nl
  [arXiv:1307.6346 [hep-ex]].

\bibitem{Evans:2013jna}
  J.~A.~Evans, Y.~Kats, D.~Shih and M.~J.~Strassler,
  %``Toward Full LHC Coverage of Natural Supersymmetry,''
  JHEP {\bf 1407} (2014) 101
  [arXiv:1310.5758 [hep-ph]].

\bibitem{Ellwanger:2004xm}
  U.~Ellwanger, J.~F.~Gunion and C.~Hugonie,
  %``NMHDECAY: A Fortran code for the Higgs masses, couplings and
  %decay widths in the NMSSM,'' 
  JHEP {\bf 0502} (2005) 066
  [hep-ph/0406215].

\bibitem{Ellwanger:2005dv}
  U.~Ellwanger and C.~Hugonie,
  %``NMHDECAY 2.0: An Updated program for sparticle masses, Higgs
  %masses, couplings and decay widths in the NMSSM,'' 
  Comput.\ Phys.\ Commun.\  {\bf 175} (2006) 290
  [hep-ph/0508022].

\bibitem{Das:2011dg}
  D.~Das, U.~Ellwanger and A.~M.~Teixeira,
  %``NMSDECAY: A Fortran Code for Supersymmetric Particle Decays in the Next-to-Minimal Supersymmetric Standard Model,''
  Comput.\ Phys.\ Commun.\  {\bf 183} (2012) 774
  [arXiv:1106.5633 [hep-ph]].

\bibitem{Djouadi:1997yw}
  A.~Djouadi, J.~Kalinowski and M.~Spira,
  %``HDECAY: A Program for Higgs boson decays in the standard model and its supersymmetric extension,''
  Comput.\ Phys.\ Commun.\  {\bf 108} (1998) 56
  [hep-ph/9704448].

\end{thebibliography}
\end{document}